\documentclass[aps,prl,showpacs,amsmath,twocolumn,amssymb,superscriptaddress,letterpaper]{revtex4}
\usepackage{graphicx,color}
\usepackage{amssymb}   % for math
\usepackage{amsmath}
\usepackage{epstopdf}
\usepackage{natbib}
\usepackage{hyperref}
\usepackage{bm}

\newcommand{\nd}{{\vphantom{\dagger}}}
\begin{document}
\title{Asymmetric Josephson Effect in Inversion Symmetry Breaking Topological Materials }
\author{ Chui-Zhen Chen }
\affiliation{Department of Physics, Hong Kong University of Science and Technology, Clear Water Bay, Hong Kong, China. }
\author{James Jun He}
\affiliation{Department of Physics, Hong Kong University of Science and Technology, Clear Water Bay, Hong Kong, China. }
\author{Mazhar Ali}
\affiliation{Max Plank Institute for Microstructure Physics, Weinberg 2, 06120 Halle, Germany}
\author{Gil-ho Lee} \thanks{lghman@postech.ac.kr}
\affiliation{ Department of Physics, POSTECH, Pohang, South Korea }
\author{K C Fong} \thanks{kc.fong@raytheon.com}
\affiliation{ Raytheon BBN Technologies, Quantum Information Processing Group, Cambridge, MA 02138, USA}
\author{K. T. Law} \thanks{phlaw@ust.hk}
\affiliation{Department of Physics, Hong Kong University of Science and Technology, Clear Water Bay, Hong Kong, China. }
\begin{abstract}
Topological materials which possess topologically protected surface states have attracted much attention in recent years. In this work, we study the critical current of superconductor/inversion symmetry breaking topological material/superconductor junctions. We found surprisingly that, in topological materials with broken inversion symmetry, the magnitude of the critical Josephson currents $|I^{+}_c(B)|$ at fixed magnetic field $B$ is not the same for critical currents $|I^{-}_c(B)|$ flowing in the opposite direction. Moreover, the critical currents violate the $| I_{c}^{\pm}(B)| = |I_{c}^{\pm}(-B)|$ relation and give rise to asymmetric Fraunhofer patterns. We call this phenomenon asymmetric Josephson effect (AJE).  AJE can be use to detect inversion symmetry breaking in topological materials such as in quantum spin Hall systems and Weyl semimetals.
\end{abstract}
\pacs{}

\maketitle

{\emph {Introduction.}}---
Over the past decade, there has been an intense interest in the study of topological materials such as topological insulators which possess surface states \cite{Qi2011,Hasan2010}. The surface states are protected by the bulk insulating gap and time-reversal symmetry. In more recent years, it was shown that protected surface states can exist even in gapless systems such as Weyl semimetals when inversion symmetry breaking splits a Dirac point into two Weyl points with opposite chirality \cite{Wan2011,Balents2011,Weng2015,SMHuang2015}. The projections of the Weyl points on the surface Brillouin zones are connected by Fermi arcs which result in conducting surface states on the surface of the Weyl semimetal. Surface Fermi arcs have been observed experimentally through angle resolved photoemission spectroscopy (ARPES) experiments \cite{Xu2015,Lv2015}. However, the transport studies of Weyl semimetals have been mostly focused on chiral anomaly \cite{DTSon,Nielsen1981,Kim2013,Xiong2015} and other bulk properties of Weyl semimetals \cite{Gooth2017,Hirschberger,Huang2015,Li2015,Sachdev}. On the other hand, electronic transport signatures of Fermi arc states of Weyl semimetals have not been well explored theoretically and experimentally \cite{Moll2016,Potter2014,Chen2017,Shvetsov}. We shall investigate the direct transport signature of inversion symmetry breaking in these topological materials.

In this work, we study the critical currents $I_c$ as a function of magnetic field $B$ in a superconductor/inversion symmetry breaking topological material/superconductor Josephson junction as depicted in Fig.1. The Josephson current of the junction is mediated by the edge states or surface states as well as the bulk states of the topological material. We show that the magnitude of critical current across the junction $|I_{c}^{+}(B)|$ does not equal to the critical current flowing in the opposite direction $|I_{c}^{-}(B)|$ at fixed magnetic field $B$ such that $|I_{c}^{+}(B)| \neq |I_{c}^{-}(B)|$. Moreover, the critical currents are different when the magnetic field switches sign, namely, $| I_{c}^{\pm}(B)| \neq |I_{c}^{\pm}(-B)|$. This gives rise to asymmetric Fraunhofer patterns. This phenomenon is in sharp contrast to conventional Josephson junctions in which $| I_{c}^{\pm}(B)| = |I_{c}^{\pm}(-B)|$ and $I_c$ is independent of the direction of the Josephson current. We call this phenomenon asymmetric Josephson effect (AJE). Particularly, we show that AJE is particularly pronounced for the edge/surface states and it can be used to detect the inversion symmetry breaking effects in Weyl semimetal through the Fermi arc induced AJE.

\begin{figure}[tbh]
\centering
\includegraphics[width=3.5in]{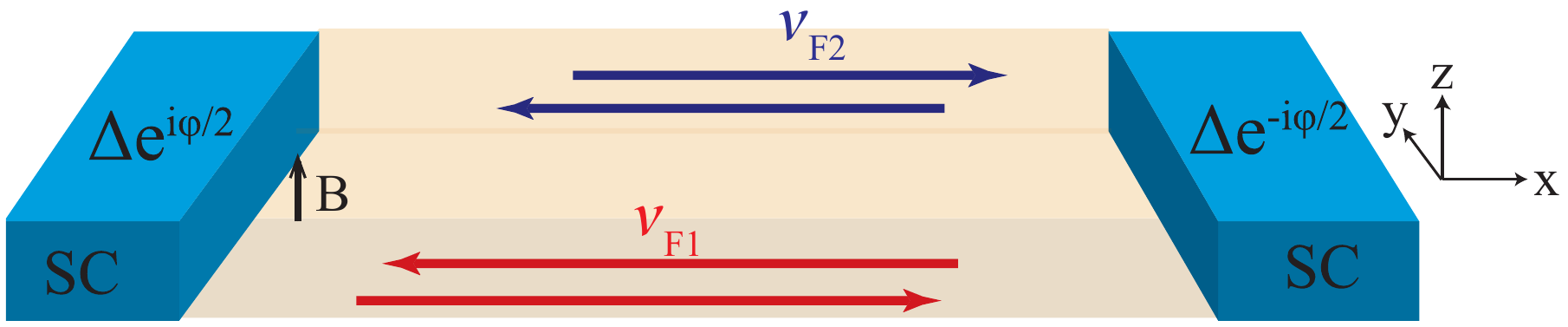}
\caption{(Color online). Schematic picture of a superconductor/topological material/superconductor junction where $\varphi$ is the phase difference between two superconductors. The edge (or surface) states on two sides of a topological material have different Fermi velocities $v_{F1}^\protect\nd$ and $v_{F2}^\protect\nd$ when inversion symmetry is broken and this can give rise to asymmetric Josephson effect (AJE) in the presence magnetic field $B$.
\label{fig1} }
\end{figure}

To understand the origin of AJE in topological materials, we first start with a Josephson junction formed by two superconductors and an inversion symmetry breaking two-dimensional (2D) topological insulator with  helical edge states. Due to inversion symmetry breaking,  the top (blue) and bottom (red)  helical edge states have different Fermi velocities as depicted in Fig.2. In the Josephson junction, the superconductors induce a superconducting gap on the edge states and create Andreev bound states. Due to different Fermi velocities, the Andreev bound state spectrums of the top and bottom edges are different as depicted in Fig.3. This results in different Josephson current contribution from the top and bottom edges. Furthermore, the magnetic field generates a phase difference between the two Josephson current channels on the two edges and results in AJE. The AJE is manifested by the asymmetric Fraunhofer pattern of the Josephson junction as shown in Fig.2c and Fig.2d.  As shown in Fig.4, we show that the Fermi arc states of Weyl semimetals can give rise to pronounced AJE.

{\emph{AJE in 2D topological insulator with broken inversion symmetry}} --- To start with, we study a superconductor/2D topological insulator/superconductor junction. The 2D topological insulator is described by a square lattice with four orbitals on each site \cite{Bernevig2006} with an additional term which breaks inversion symmetry:
\begin{eqnarray}
  H_{TI}^\nd&=& \sum_{k_x,k_y}\psi_{\bf k}^{\dagger} \{\Gamma\sin k_x \tau_0\sigma_3 +  M_{\bf k} \tau_3\sigma_0 +  A \sin k_x \tau_1\sigma_3 \nonumber\\
    & &+A \sin k_y \tau_2\sigma_0 \}\psi_{\bf k}^\nd,
    \label{eq1}
\end{eqnarray}
where the Pauli matrices $\tau_{1,2,3}$ ($\sigma_{3}$) and the unit matrix $\tau_0$($\sigma_{0}$) are defined in the orbital (spin) space,
and $\psi_{\bf k}^{\dagger}$ is a four component fermionic operator.
$M_{\bf k}=m_{0} + 2m_1 (2-\cos k_x  -\cos k_y)$ determines the energy gap of the system with the momentum $k_{x,y}$ and $A$ couples two orbitals. When the $\Gamma$ term is non-zero, the inversion symmetry is broken.
%in the first Brillouin zone of an $L_x \times L_y $ square lattice with the lattice constant $a \equiv 1$.

Fig.2a and Fig.2b depict the energy spectrums of the system in the topological regime with open boundary condition in the y-direction and there are helical edge states propagating at the edge of the sample. The top edges (blue lines) and bottom edges (red lines) have the same Fermi velocity in the presence of the inversion symmetry (Fig.2a), while the Fermi velocities are different if the inversion symmetry is broken with $\Gamma=0.5$ (Fig.2b).
This topological insulator can form a Josephson junction with two superconductors as shown in Fig.1, where $\Delta e^{\pm i\varphi/2}$ denotes the pairing order parameter of the superconductors.
Fig.2c and Fig.2d depict the critical Josephson current as a function of magnetic flux $\Phi$ at the temperature $T = \Delta/30$  by recursive Green's function \cite{Asano2003,Furusaki}. Here, we assume that fermion parity is not conserved at the junction so that $4\pi$ Josephson effect is absent and this is consistent with the results of a recent experiment \cite{Molenkamp2017}.

\begin{figure}[tbh]
\centering
\includegraphics[width=3.4in]{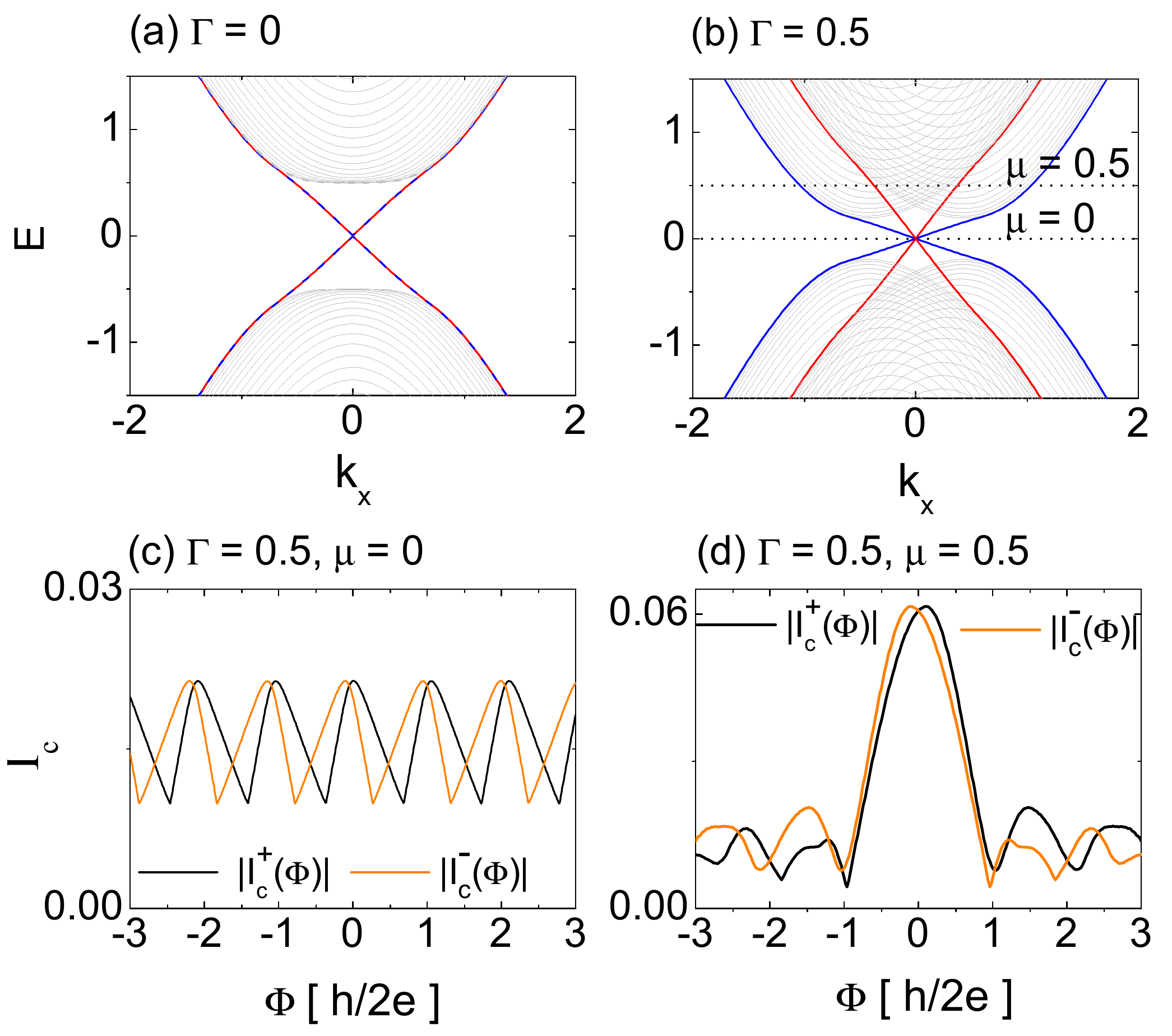}
\caption{(Color online). (a) and (b)  Energy spectrums of a two-dimensional topological insulator with $\Gamma=0$ and $\Gamma=0.5$, respectivley. When $\Gamma$ is finite, the top (blue line) and bottom (red line) edges acquire different Fermi velocities $v_{F1}^\protect\nd$ and $v_{F2}^\protect\nd$, respectively. In (c) an (d), superconductors are attached to the topological insulator to form a Josephson junction as shown in Fig.1. The pairing potential is $\Delta=0.05$. (c) and (d) depict the critical current $I_c^{\pm}(\Phi)$ as a function of magnetic flux $\Phi[h/2e]$ of the Josephson junction for $\Gamma=0.5$ at chemical potential $\mu=0$ and $\mu=0.5$ respectively. As shown in (b), at $\mu=0$, only the edge states contribute to the transport. It is clearly shown in (c) that $|I_{c}^{+}(\Phi)| \neq |I_{c}^{-}(\Phi)|$ and $|I_{c}^{+}(\Phi)| \neq |I_{c}^{+}(-\Phi)|$. (d) At $\mu=0.5$, the bulk states also contribute to the bulk transport. AJE is more pronounced when edge states dominate the transport.
\label{fig2} }
\end{figure}

In the presence of inversion symmetry, the critical Josephson currents are the same in opposite directions, namely, $| I^{+}_{c}(\Phi)| = |I^{-}_{c}(\Phi)|$.
Surprisingly, when inversion symmetry is broken as shown in Fig.2c and Fig.2d, the critical current across the junction $I_{c}^{+}$ does not equal to the critical current flowing in the opposite direction $I_{c}^{-}$ at fixed magnetic field such that $|I_{c}^{+}(\Phi)| \neq |I_{c}^{-}(\Phi)|$. Moreover, the critical currents also manifests asymmetric Fraunhofer patterns as depicted in Fig.2c and Fig.2d such that $| I_{c}^{\pm}(\Phi)| \neq |I_{c}^{\pm}(-\Phi)|$. This is in sharp contrast to conventional Josephson effect and we call this phenomenon asymmetric Josephson effect. It is important to note that the directional dependence of the critical current $|I_{c}^{+}(\Phi)| \neq |I_{c}^{-}(\Phi)|$ and the asymmetric Fraunhofer pattern are connected to each other. Due to time-reversal invariant, we have the condition that the critical current is unchanged when both the direction of the current and the direction of the magnetic field are changed, namely, $ |I_c^{+}(\Phi)| = |I_c^{-}(-\Phi)|$. As a result, $|I_{c}^{+}(\Phi)| \neq |I_{c}^{-}(\Phi)|$ implies $| I_{c}^{\pm}(\Phi)| \neq |I_{c}^{\pm}(-\Phi)|$. It is important to note that asymmetric Fraunhofer pattern similar to Fig.2d has been observed recently \cite{Molenkamp2017}, but the origin of the effect was not known. In this work, we provide an explanation of the asymmetric Fraunhofer pattern.

\begin{figure}[tbh]
\centering
\includegraphics[width=3.4in]{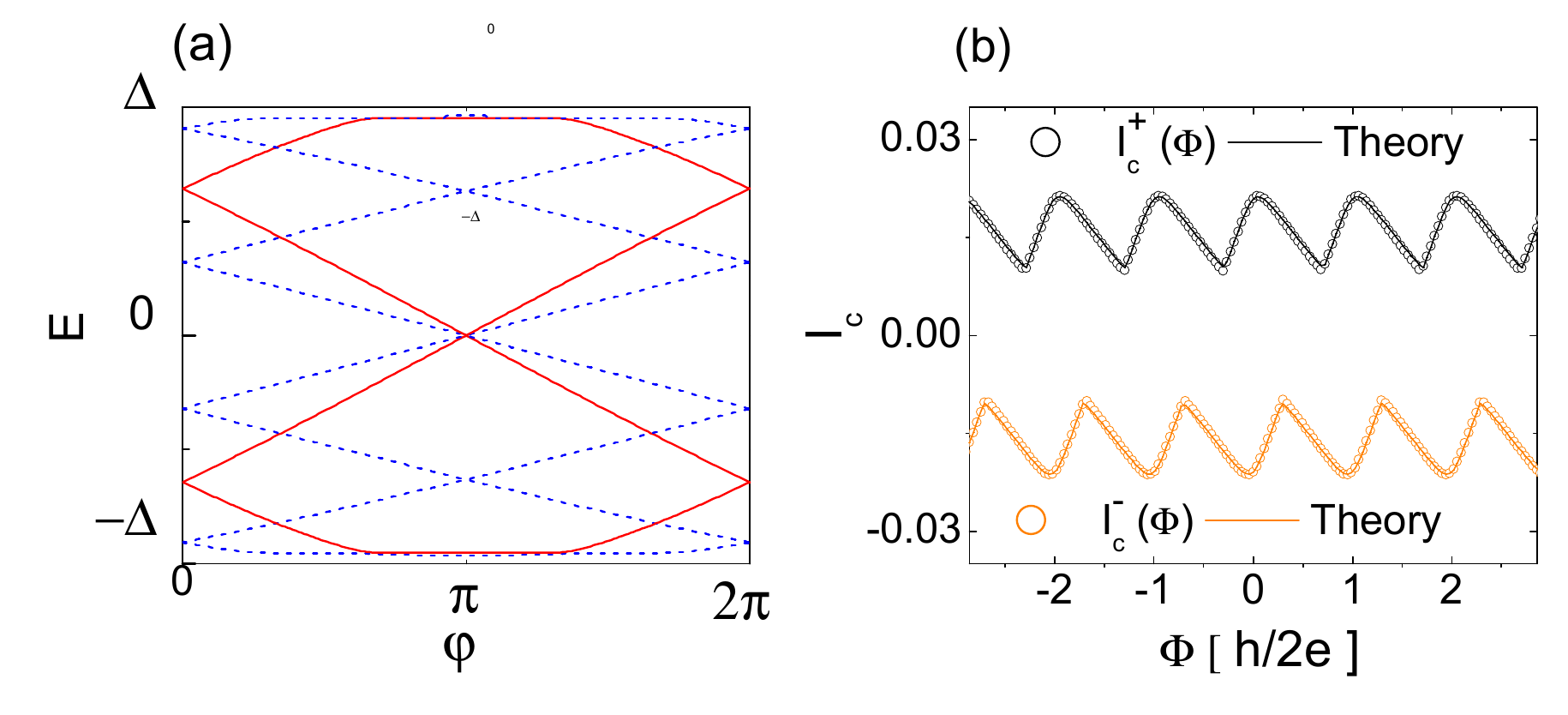}
\caption{(Color online).  (a) Andreev bound state spectrum of the superconductor/2D topological insulator/superconductor junction as a function of $\varphi$ when inversion-symmetry is broken with $\Gamma=0.5$.
The dash solid blue lines  and red lines denote the energy spectrum of upper and lower edges, respectively. The superconducting paring potential $\Delta=0.05$ and other parameters are the same as those of Fig.\ref{fig2}b. (b) The numerical results of Fig.2c, which shows the AJE, can be easily reproduced using the phenomenological theory using different $I_{1n}$ and $I_{2n}$ in the presence of magnetic field.
\label{fig3} }
\end{figure}

To understand the origin of the AJE, we investigate the energy spectrum of Andreev bound states in the Josephson junction. In the presence of inversion symmetry, the energy spectrum has two fold degeneracies, since the top edges and bottom edges have exactly the same spectrum. On the other hand, the energy spectrum of Andreev bound states from the top edges and bottom edges are different in Fig\ref{fig3}a, because the two edge states have different Fermi velocities when the inversion symmetry is broken. As a result, the supercurrent from the top edges $I_1$ and the bottom edges $I_2$ are different due to the different Andreev bound state spectrums of the two edges. This gives rise to AJE as discussed below.

{\emph{Phenomenological Theory}} -- In general, the total Josephson current $I(\Phi,\varphi)$ carried by the two edges states can be described as \cite{Asano2003}
\begin{eqnarray*}
  I(\Phi,\varphi)&=& \sum_{n=1}^{m} I_{1n} \sin(n \varphi+ n\Phi)+ I_{2n} \sin(n \varphi-n\Phi).
\end{eqnarray*}
Here, $I_{ln}$ indicates the Josephson current carried by the $l$ edge at $n$th order, $\Phi$ represents the magnetic phase in the normal region and $\varphi$ is  the phase difference between two s-wave superconductors. If the top edge current $I_1$ is the same as the bottom edge current $I_2$ due to inversion symmetry, we have $I_{1n}=I_{2n}$ and the Josephson current can be written as $I(\Phi,\varphi) = \sum_{n=1}^{m} (I_{1n} + I_{2n} )\cos(n\Phi)\sin(n \varphi)$. This implies that the Josephson current is always symmetric with respect to the signs of the magnetic field, namely, $I(\Phi,\varphi)=I(-\Phi,\varphi)$.
In general, the top and bottom edges can have different energy spectrums due to inversion symmetry breaking as discussed above. Therefore, the two sets of coefficients $I_{1n}$ and $I_{2n}$ can be different.

%\begin{figure}[tbh]
%\centering
%\includegraphics[width=3.4in]{Fig4.pdf}
%\caption{(Color online). Critical Josephson current $I_c^{\pm}$  versus magnetic flux $\Phi[h/2e]$
 %for topological insulator at (a) $\Gamma = 0$ and (b) $\Gamma = 0.5$ with $\mu = 0.15$.
 %The numerical data (open circles) are the same as those of Fig.2(c) and (d),
 %which can be well fitted by phenomenological theory (solid lines).
%\label{fig4} }
%\end{figure}

%\begin{figure}[tbh]
%\centering
%\includegraphics[width=3.4in]{Fig5.pdf}
%\caption{(Color online). (a) and (b) plot the energy spectrum of $L_z=2$ and $L_z= 10$ layers Weyl semimetal thin film, %respectively.
%(c) and (d) show critical Josephson current $I_{c}^{\pm}$ versus magnetic flux $\Phi[h/2e]$
%of  $L_z=2$ and $L_z= 10$ layers Weyl semimetal thin film, respectively.
%The dash lines $I_{c}^{\pm}(-\Phi)$ are time-reversal plot of solid lines $I_{c}^{\pm}(\Phi)$.
%The parameters are $\Delta=0.05$, m=-0.2 and $t_z=1.5$ with other model parameters the same as those of Fig.\ref{fig2}(b).
%\label{fig5} }
%\end{figure}

In Fig.\ref{fig3}b, we show that  the numerical results  of the Josephson current $I_c^{\pm}$ as a function of flux $\Phi$ can be well fitted to phenomenological theory (solid lines).
The asymmetric critical Josephson currents are fitted by the $I_{1n}\neq I_{2n}$ and the Josephson currents of two edge states have the different magnitudes,
which results from the asymmetric Fermi velocity of the two edge states $v_{F1}^{\nd}\neq v_{F2}^{\nd}$ as we discussed above.
The consistence between the phenomenological theory and numerical results demonstrates that the AJE originates from
the different Fermi velocities of the two sets of edge modes.
%If we further include a small amount bulk states, the Fraunhofer pattern will be tilted as in

\begin{figure}[tbh]
\centering
\includegraphics[width=3.4in]{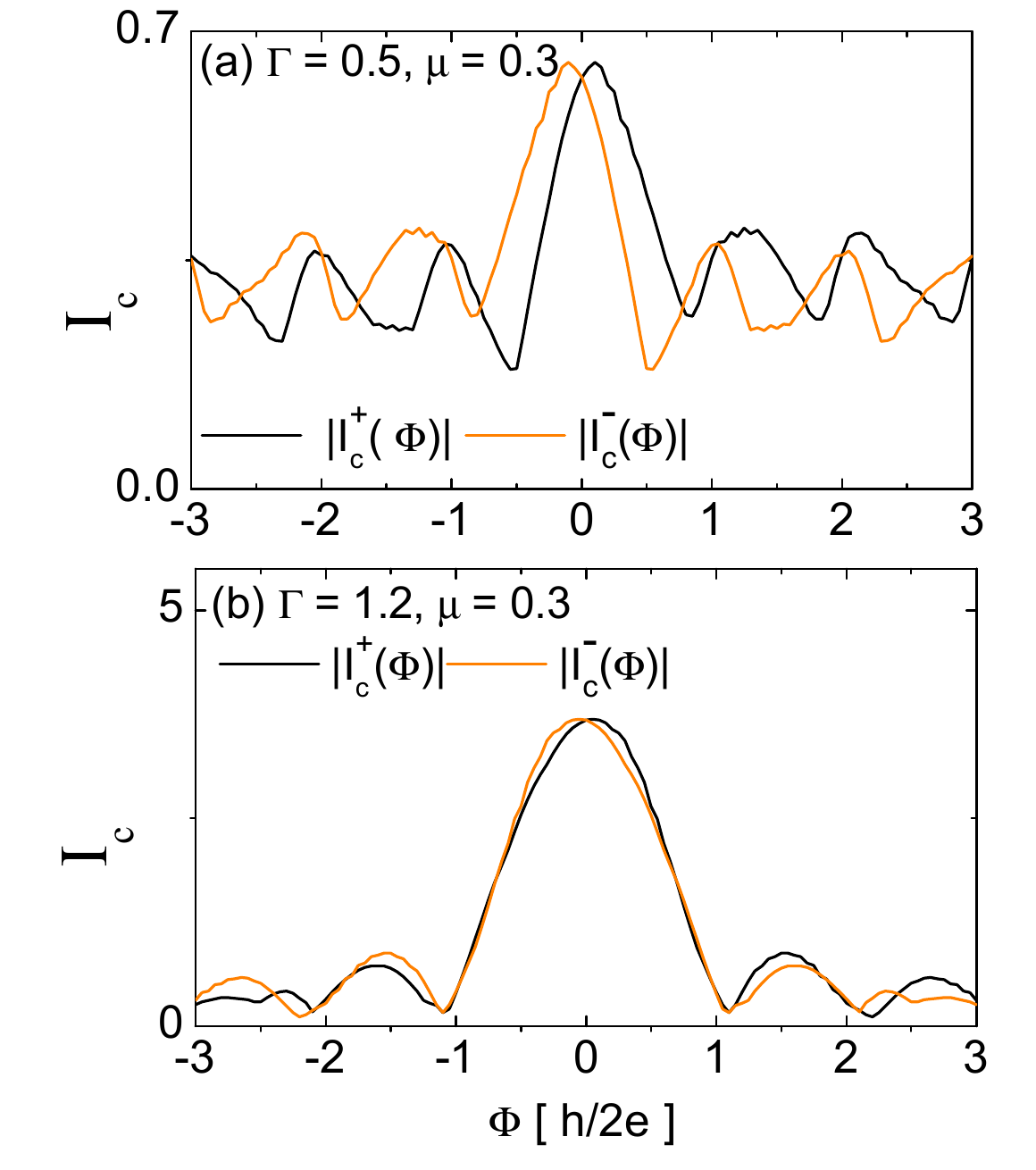}
\caption{(Color online).
(a) and (b) show critical Josephson current $I_{c}^{\pm}$ versus magnetic flux $\Phi[h/2e]$
for type-I Weyl semimetal $\Gamma = 0.5$ and  type-II Weyl semimetal $\Gamma = 1.2$, respectively.
The Josephson current is mediated by both the Fermi arcs on the surface and the Weyl nodes in the bulk,
because of $I_{c}^{\pm}$ has a central peak.
AJE is more pronounced when the Fermi arc surface states dominate the transport.
The parameters are $\Delta=0.05$, m=-0.2 and $t_z=1.5$.
\label{fig4} }
\end{figure}

{\emph{ AJE in Weyl semimetals}} -- Next we show that the Fermi arc states of Weyl semimetal can give rise to pronounced AJE in the  a superconductor/Weyl semimetal/superconductor junction.
Stacking the two-dimensional TI Hamiltonian $H_{TI}^\nd$ in z direction, we obtain a three-dimensional  Weyl semimetal Hamiltonian \cite{Kourtis2016,Gilbert}
\begin{eqnarray}
  \mathcal{H}&=& \sum_{\bf k}\psi_{\bf k}^{\dagger} \{ \Gamma\sin k_x \tau_0\sigma_3+ (M_{\bf k} - t_z\cos k_z )\tau_3\sigma_0 \nonumber\\
    & & +  A \sin k_x \tau_1\sigma_3 +A\sin k_y \tau_2\sigma_0 \}\psi_{\bf k}^\nd
    \label{eq2}
\end{eqnarray}
where the Pauli matrices $\tau_{1,2,3}$ ($\sigma_{3}$) and the unit matrix $\tau_0$($\sigma_{0}$) are defined in the orbital (spin) space, and  $\psi_{\bf k}^{\dagger}$ is a four component fermionic operator with momentum ${\bf k}$.
$t_z=1.5$ is the hopping energy in z direction and $A$ couples two orbitals in x-y plane.
Here  $M_{\bf k}=m_0 + 2m_1 (2-\cos k_x  -\cos k_y)$ determines the position of Weyl nodes with $m_0=-0.2$.
The Weyl nodes are located at (0,0,$\pm\arccos(m_0/t_z)$),
while the inversion symmetry breaking $\Gamma$ term creates a tilted effect to the Weyl nodes \cite{Gilbert}.
The Weyl nodes are type-I for $\Gamma<A$ and become type-II when $\Gamma>A$  \cite{Gilbert,Soluyanov2015}.
In general, the projections of the Weyl points on the surface Brillouin zones are connected by Fermi arcs, which result in conducting surface states on the surface of the Weyl semimetal.
Because these surface arc states on two opposite surfaces of the Weyl semimetal can have different Fermi velocities when inversion symmetry is broken by $\Gamma$. This will give rise to the AJE in the superconductor/Weyl semimetal/superconductor junction in the following.

%Figs.\ref{fig5}(a) and \ref{fig5}(b) plot the energy spectrums of the Weyl semimetal thin film.
%The system has a bulk gap due to size effect in z direction [see Fig.\ref{fig5}(a)], which can be regard as alternative stacking %of  a few layer TIs  and normal insulators.
%Due to the broken inversion symmetry, the top and bottom edge modes have different Fermi velocities as we discussed above.
%Increasing the Weyl semimetal thin film to $N_z=10$ layers, the bulk gap becomes very small and thus the edge states generally %coexist with the bulks states.
%If we attach the Weyl semimetal thin film to two superconducting leads with different phases,
%we find the AJE in Fig.\ref{fig5}(c), namely,  the critical current across the junction $I_{c}^{+}$ does not equal to the %critical current flowing in opposite direction $I_{c}^{-}$ at fixed magnetic field  such that $I_{c}^{+}(\Phi) \neq %I_{c}^{-}(\Phi)$. Moreover, the AJE can result in the asymmetric Fraunhofer pattern i.e. $| I_{c}^{\pm}(\Phi)| \neq %|I_{c}^{\pm}(-\Phi)|$.
%In Fig.\ref{fig5}(d), there is a large asymmetric central peak in the Fraunhofer pattern, imply that the AJE still survives when %the Josephson current are carried by the edge states as well as the bulk states.
In Fig.\ref{fig4}, the Josephson current is mediated by both the Fermi arc states on the surface as well as the states near the Weyl nodes in the bulk.
In this case, we can find pronounced AJE for type-I Weyl semimetal in Fig.\ref{fig4}a, namely,  the critical current across the junction $I_{c}^{+}$ does not equal to the critical current flowing in the opposite direction $I_{c}^{-}$ at fixed magnetic field  such that $|I_{c}^{+}(\Phi)| \neq |I_{c}^{-}(\Phi)|$. Moreover, the AJE is also manifested by the asymmetric Fraunhofer pattern in which $| I_{c}^{\pm}(\Phi)| \neq |I_{c}^{\pm}(-\Phi)|$ as shown in Fig.\ref{fig4}a.
In Fig.\ref{fig4}b, the AJE is also observed for type-II Weyl semimetal, even though there are more Josephson currents carried by the bulk states in type-II Weyl semimetals.
Therefore, we conclude that AJE is more pronounced for surface state dominated transport and it can be used as a transport signature for broken inversion symmetry ( or surface and edge states with differing Fermi velocities ) in topological insulators and Weyl semimetals.

{\emph{Discussion and conclusion}--
In summary, we uncovered an unusual Josephson effect in the inversion symmetry breaking topological materials. It is found that the  magnitude of critical Josephson current across the junction $I_{c}^{+}$ does not equal to the critical Josephson current flowing in opposite direction $I_{c}^{-}$ at fixed magnetic field B such that $|I_{c}^{+}(B)| \neq |I_{c}^{-}(B)|$.
We call this phenomenon AJE. This can give rise to asymmetric Fraunhofer patterns which violate $| I_{c}^{\pm}(B)| = |I_{c}^{\pm}(-B)|$. This is in sharp contrast to conventional Josephson junctions. The AJE is a very general phenomenon, which can originate from topologically nontrivial or trivial surface states with differing Fermi velocities.
We emphasize that the AJE discussed here is an intrinsic property of the system, which is disctinct from the asymmetric Fraunhofer patterns induced by an external in-plane magnetic field with in-plane component  \cite{Suominen}.
Interestingly, the AJE shown in Fig.\ref{fig2}d qualitatively agree with recently observed asymmetric Fraunhofer pattern in  superconductor/quantum spin Hall insulator/superconductor Josephson junction \cite{Molenkamp2017}.
They found that the critical current violates $|I_c^{+}(B)|= |I_c^{-}(B)|$ but follows the symmetry relation $|I_c^{+}(B)|=|I_c^{-}(-B)|$, which is the same as what we discussed here.

{\emph{Acknowledgement.}}--- We thank Patrick A Lee, Benjamin T Zhou and Dong-Hui Xu for illuminating discussions. The authors thank the support of HKRGC and Croucher Foundation through HKUST3/CRF/13G, 602813, 605512, 16303014 and Croucher Innovation Grant. KCF thanks the funding support from Raytheon BBN Technologies.


\begin{thebibliography}{WSM_JJ}
\bibitem{Hasan2010}M. Z. Hasan and C. L. Kane, Rev. Mod. Phys. {\bf82}, 3045 (2010).
\bibitem{Qi2011} X.-L. Qi and S.-C. Zhang, Rev. Mod. Phys. {\bf83}, 1057 (2011).
\bibitem{Wan2011}  X. Wan et al., Phys. Rev. B {\bf83}, 205101 (2011).
\bibitem{Balents2011} A. A. Burkov and L. Balents, Phys. Rev. Lett. {\bf 107}, 127205 (2011).


\bibitem{Weng2015} H. Weng, C. Fang, Z. Fang, B. A. Bernevig, and X. Dai, Phys. Rev. X {\bf 5}, 011029 (2015).
\bibitem{SMHuang2015} S. M. Huang, S. Y. Xu, I. Belopolski, C. C. Lee, G. Q. Chang, B. K. Wang, N. Alidoust, G. Bian, M. Neupane, A. Bansil, H. Lin, and M. Z. Hasan, Nat. Commun. {\bf 6}, 7373 (2015).
\bibitem{Xu2015}S. Y. Xu, C. Liu, S. K. Kushwaha, R. Sankar, J. W. Krizan, I. Belopolski, M. Neupane, G. Bian, N. Alidoust, T. R. Chang, H. T. Jeng, C. Y. Huang, W. F. Tsai, H. Lin, P. P. Shibayev, F. C. Chou, R. J. Cava, and M. Z. Hasan, Science {\bf 347}, 294 (2015).
\bibitem{Lv2015} B. Q. Lv, H. M. Weng, B. B. Fu, X. P. Wang, H. Miao, J. Ma, P. Richard, X. C. Huang, L. X. Zhao, G. F. Chen, Z. Fang, X. Dai, T. Qian, and H. Ding, Phys. Rev. X {\bf 5}, 031013 (2015).
\bibitem{Nielsen1981} H. Nielsen and M. Ninomiya, Phys. Lett. B {\bf130}, 389(1981).
\bibitem{DTSon} D. T. Son and B. Z. Spivak, Phys. Rev. B {\bf 88}, 104412 (2013).
\bibitem{Kim2013} H.-J. Kim, K.-S. Kim, J.-F. Wang, M. Sasaki, N. Satoh, A. Ohnishi, M. Kitaura, M. Yang, and L. Li, Phys. Rev. Lett. {\bf 111}, 246603 (2013).
\bibitem{Xiong2015} J. Xiong, S. K. Kushwaha, T. Liang, J. W. Krizan, M. Hirschberger, W. Wang, R. J. Cava, and N. P. Ong, Science {\bf350}, 413 (2015).
\bibitem{Huang2015}X. C. Huang et al., Phys. Rev. X {\bf5}, 031023 (2015).
\bibitem{Li2015}C. Z. Li, L. X. Wang, H. W. Liu, J. Wang, Z. M. Liao, and D. P. Yu, Nat. Commun. 6, 10137 (2015).
\bibitem{Sachdev} A. Lucas, R. A. Davison and and S. Sachdev, Proc. Natl Acad. Sci. USA {\bf113}, 9463  (2016).
\bibitem{Hirschberger} M. Hirschberger, S. Kushwaha, Z. Wang, Q. Gibson, C. A. Belvin, B. A. Bernevig, R. J. Cava, and N. P. Ong, Nat. Mater. {\bf15}, 1161 (2016).
\bibitem{Gooth2017}J. Gooth et al.  Nature (London)  {\bf547}, 324 (2017).
\bibitem{Potter2014} A. C. Potter, I. Kimchi, and A. Vishwanath, Nat. Commun. {\bf5}, 5161 (2014).
\bibitem{Moll2016} P. J. Moll, N. L. Nair, T. Helm, A. C. Potter, I. Kimchi, A. Vishwanath, and J. G. Analytis, Nature (London) {\bf535}, 266 (2016).
\bibitem{Chen2017}A. Chen, D. I. Pikulin, and M. Franz, Phys. Rev. B {\bf95}, 174505 (2017).
\bibitem{Shvetsov} O. O. Shvetsov, A. Kononov, A. V. Timonina, N. N. Kolesnikov, and E. V. Deviatov,
arXiv:1801.09551.
%\bibitem{Kaminski2015} L. Huang, T. M. McCormick, M. Ochi, Z. Zhao, M.-T. Suzuki, R. Arita, Yun Wu, D. Mou, H. Cao, J. Yan, N. Trivedi, and A. Kaminski, Nat. Mater. {\bf 15}, 1155 (2016).
%\bibitem{SZhou2016} K. Deng et.al., Nat. Phys. {\bf 12}, 1105 (2016).
\bibitem{Bernevig2006}B. A. Bernevig, T. L. Hughes, and S.-C. Zhang, Science {\bf314}, 1757 (2006).


\bibitem{Furusaki} A. Furusaki, Physica (Amsterdam) {\bf 203B}, 214 (1994); Y. Asano, Phys. Rev. B {\bf63}, 052512 (2001).
\bibitem{Asano2003} Y. Asano, Y. Tanaka, M. Sigrist, and S. Kashiwaya, Phys. Rev. B {\bf 67}, 184505 (2003).
\bibitem{Molenkamp2017} E. Bocquillon, R. S. Deacon, J. Wiedenmann, P. Leubner, T. M. Klapwijk, C. Br\"{u}ne, K. Ishibashi, H. Buhmann, and L. W. Molenkamp, Nat. Nanotechnol. {\bf12}, 137 (2017).
\bibitem{note1} In the fitting of Fig.\ref{fig3}, the magnetic flux is redefine as $\Phi=A_{eff} \times B$  to include the finite broadening of the edge states, where $B$ is magnetic field  and $A_{eff}=L\times W/1.05$ is effective area with the length $L$ and the width $W$.

\bibitem{Kourtis2016} S. Kourtis, J. Li, Z. Wang, A. Yazdani, and B. A. Bernevig, Phys. Rev. B {\bf 93}, 041109 (2016).
\bibitem{Gilbert} M. J. Park, B. Basa, and M. J. Gilbert, Phys. Rev. B {\bf95}, 094201 (2017).
%\bibitem{Trivedi2017} T. M. McCormick, I. Kimchi, and N. Trivedi, Phys. Rev. B {\bf 95}, 075133 (2017).
\bibitem{Soluyanov2015}A. A. Soluyanov, D. Gresch, Z. Wang, Q. Wu, M. Troyer, X. Dai, and B. A. Bernevig, Nature (London) {\bf 527}, 495 (2015).
%\bibitem{ZWang2016}Z. Wang, D. Gresch, A. A. Soluyanov, W. Xie, S. Kushwaha, X. Dai, M. Troyer, R. J. Cava, and B. A. Bernevig, Phys. Rev. Lett. {\bf 117}, 056805 (2016).
\bibitem{Kaminski2015} L. Huang, T. M. McCormick, M. Ochi, Z. Zhao, M.-T. Suzuki, R. Arita, Yun Wu, D. Mou, H. Cao, J. Yan, N. Trivedi, and A. Kaminski, Nat. Mater. {\bf 15}, 1155 (2016).
\bibitem{SZhou2016} K. Deng, G. Wan, P. Deng, K. Zhang, S. Ding, E. Wang, M. Yan, H. Huang, H. Zhang, Z. Xu, J. Denlinger, A. Fedorov, H. Yang, W. Duan, H. Yao, Y. Wu, S. Fan, H. Zhang, X. Chen, and S. Zhou, Nat. Phys. {\bf 12}, 1105 (2016).
\bibitem{Suominen} H. Suominen, J. Danon, M. Kjaergaard, K. Flensberg, J. Shabani, C. Palmstrøm, F. Nichele, and C. Marcus, Phys. Rev. B {\bf95}, 035307 (2017).
\end{thebibliography}
\end{document}